\documentclass[conference]{IEEEtran}
\usepackage{graphicx}
\usepackage[cmex10]{amsmath}
\usepackage{algorithmic}
\usepackage{array}
\usepackage[tight,footnotesize]{subfigure}
\usepackage{fixltx2e}
\usepackage{url}

\begin{document}

\title{Performance of a reconciliation method operating on a discreet quantum key distribution system}

\author{\IEEEauthorblockN{N. Benletaief, H. Rezig *Members IEEE, A. Bouallegue *Members IEEE}
\IEEEauthorblockA{Communication System laboratory Sys'Com \\National Engineering School of Tunis\\ BP 37, 1002 Tunis  Belvédère, Tunisia\\
Emails: benletaief.nedra@gmail.com, houria.rezig@enit.rnu.tn, ammar.bouallegue@enit.rnu.tn}}

\maketitle

\begin{abstract}
Reconciliation is a mechanism allowing to weed out the discrepancies between two correlated variables. It has great role in every Quantum Key Distribution protocol where the key has to be transmitted through a noisy channel or as in our case of study in presence of an eavesdropping.
In this paper, we show that for discrete-variable QKD protocols, this problem can be advantageously solved with Turbo codes. In particular, we demonstrate that our method leads to a significant improvement of Bit Error Rate, may divide it by three in presence of a eavesdropper even with great eavesdropping capability.

\end{abstract}

\section{Introduction}

Quantum Key Distribution achieves a secret random string namely the key known only to the two parties who are executing the protocol. The first protocol for quantum cryptography was proposed in 1984 by  Bennett of IBM and Brassard of the
University of Montreal and it is called the BB84 protocol~\cite{Bennett}. \\Protocol BB84 like any other protocol of quantum cryptography is based on two principal phases: a quantum phase via a one-way physical quantum channel and a public phase using an authenticated two-way classic ideal channel.
\\As categorized in~\cite{26}, a QKD protocol can usually be divided into four steps that will be illustrated below.

\begin{enumerate}
 \item Quantum transmission and reception: In order to well explain this step, we use the language of spin $\frac{1}{2}$ as the first experimental demonstration of the protocol  in 1991 used the polarization states of single photons to transmit a random key.  The protocol uses four quantum states that constitute two bases. A basis is chosen to distinguish the two values 0 and 1 without ambiguity. One choice is the rectilinear basis $\bigoplus$ where photons are polarized at angle $0^{\circ}$ ($\leftrightarrow$) or $90^{\circ}$ ($\updownarrow$) representing 0 and 1 respectively. Another choice is the diagonal basis $\bigotimes$ where 0 is represented by photons polarized at $45^{\circ}$ ($\nearrow$) and 1 by photons polarized at $135^{\circ}$ ($\nwarrow$). \\First, Alice sends quantum states chosen at random among the four states to Bob on the quantum channel. Quantum states could be sent all at once or for practical reasons  one after the other with  one restriction on the ability of establishing a one-to-one correspondence between the transmitted and the received. Next, Bob measures these incoming states in one of the two bases, chosen at random by using an independent random-number generator from that of Alice. On receiving the state, Bob informs Alice via public channels of the basis used to accept each bit. After that, Alice informs Bob, which bits were applied the correct basis. At this point, whenever they use the same basis, they have a great probability to have correlated results. However, whenever they use different bases, they get uncorrelated results. So, they discard the incorrect ones and use the remaining.
     \begin{figure}[h]
    \centering
    \includegraphics[width=9cm,height=6cm]{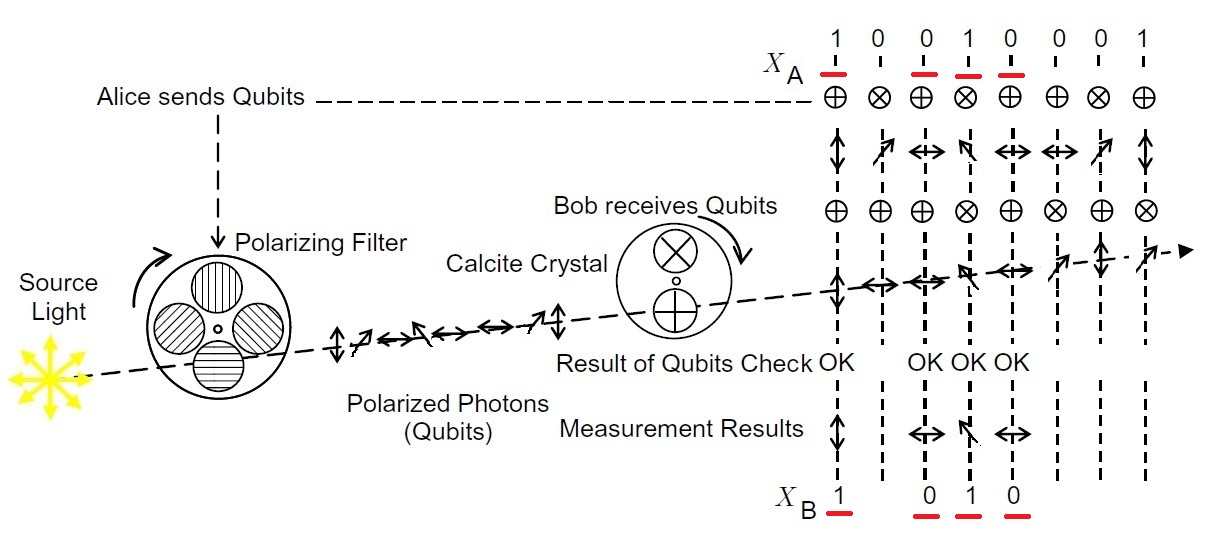}
    \caption{Steps of BB84 protocol~\cite{fig}.\label{Fig:BB84}}
    \end{figure}
     \\This process gives the two parties correlated random variables, called sifted key $X_{A}$ and $X_{B}$ (Fig.~\ref{Fig:BB84}).
  \item Channel parameter estimation: In this step, Alice and Bob estimate the channel parameters between them from announced data, in particular the joint probability distribution among Alice, Bob and Eve.
\item Information reconciliation: Theoretically unconditionally secure because it is based on the quantum laws of physics, rather than the assumed computational complexity of mathematical problems~\cite{fig}, BB84 protocol performance can be reduced by a various errors and information leakage such as limited intrinsic efficiency of the protocol, imperfect devices and eavesdropping. This step aims to improve BB84 performances by conversation over the public channel.
\item Privacy amplification: Alice and Bob shorten their bits. The resulting bits are almost statistically independent of all the information possessed by Eve.
\end{enumerate}
In our work, we are interested in correcting errors in the sifted keys. It appears clearly that reconciliation should be performed.The construction of an efficient reconciliation scheme and its validation by testing its performance are a difficult problem. Recently, Turbo codes have shown to be good candidates for this reconciliation application~\cite{nedra}.
\\This manuscript is organized as follows: In Section 2, we present an
overview of previous related works. In Section 3, we introduce the basic knowledge of Turbo codes.  In Section 4, we describe our proposed reconciliation method. In section 5, we simulate the efficiency and the correcting ability of our method  and give some results. Some conclusions and perspectives are consequently drawn in the sixth section.
\section{Previous works}
Several methods of error reconciliation for quantum cryptography have already been reported in the literature. In what follows, we will only concentrate on the most widely used ones: the best known Cascade algorithm~\cite{cascade}, the Binary algorithm~\cite{binary}, the Winnow algorithm~\cite{winnow}, and the Low-Density Parity-Check code LDPC~\cite{LDPC}.
\\In 1992, Bennett et $al$.~\cite{binary} proposed a simple and easy reconciliation method called Binary. In 1993, Brassard et $al$.~\cite{cascade} proposed a stronger ability of error correction method called Cascade. Cascade and Binary remove a single error and don't introduce additional errors to multiple errors block, instead of the Winnow, introduced by Buttler et $al$.~\cite{winnow}, algorithm because the Hamming algorithm only
reveals one single error in each block.  Cascade is an efficient method of reconciliation but its main handicap is the time of communication that it is proportional to the length of the key while for the winnow algorithm, communication time only depends on the error rate. Recently, the LDPC codes  have shown the ability to correct the same range of errors as Cascade and has the advantage to improve the safety of the used protocol.
This has motivated our approach to study Turbo codes which were proven to perform better when compared with LDPC in the literature.
\\In the context of continuous variables, modern coding techniques have been used such as Turbo codes in~\cite{turbo_con} and LDPC codes in~\cite{11} and~\cite{12}. In contrast with continous variable information reconciliation, not much has been done to adapt modern coding techniques to the discrete case. Here, it should be noted that forward error correction has the advantage of being very well known and even attaining the Shannon limit for some channels.
 This shows again that reconciliation can be advantageously solved with Turbo codes. We will describe it in more details in the next section.
\section{An overview of turbo codes}

\begin{figure}[h]
    \centering
    \includegraphics[width=9cm,height=3.5cm]{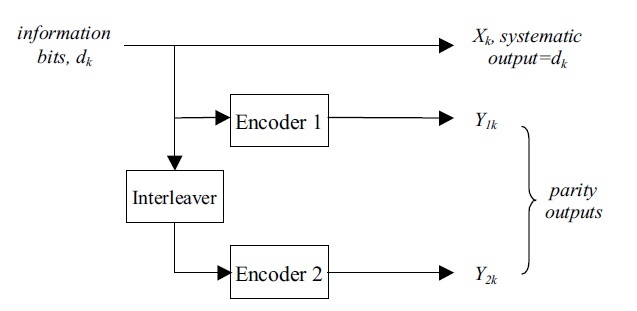}
 \caption{The generic Turbo encoder.\label{Fig:turbo}}
\end{figure}

\begin{figure}[h]
    \centering
    \includegraphics[width=9cm,height=4cm]{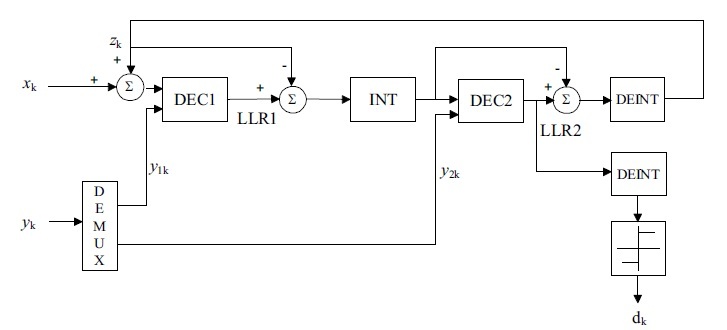}
 \caption{The generic Turbo decoder.\label{Fig:turbodec}}
\end{figure}

We start with a presentation of the Turbo code theoretical tool. Turbo coding was first introduced in 1993 by Berrou and $al$.~\cite{turbo}. The codes are constructed by using two or more component codes on different interleaved versions
of the same information sequence. The constituent codes are usually two identical recursive systematic convolutional codes (see Fig.~\ref{Fig:turbo}).
\\The input sequence to be encoded is divided into blocks of length $N$. Each block is encoded by the first encoder and interleaved before passing through the second encoder. Researchers have proved that Turbo codes can approach the Shannon limit closer than any other known forward error correcting code. The efficiency of the Turbo codes is due to the use of an iterative process at the decoder side and the presence of an interleaver at the encoder side, which adds randomness-like effect to the code.
\\The turbo decoder consists of two, or more, Soft-In Soft-Out (SISO) maximum likelihood decoders (see Fig.~\ref{Fig:turbodec}).
\\For conventional codes, the final step at the decoder yields hard-decision decoded bits or decoded symbols. Whereas, for a concatenated scheme as its the case of Turbo code, the decoding algorithm don't pass hard decisions among the decoders. But, to best exploit the information learned from each decoder, the decoding algorithm effects an exchange of soft decisions. Decoders are operating in parallel, exchanging iteratively extrinsic information. In other words, for a system with two component codes, the concept behind Turbo decoding is to pass soft decisions from the output of one decoder to the input of the other decoder, and to iterate this process several times so as to produce more reliable decisions.
\\Two families of decoding algorithms are commonly used in Turbo decoding: Soft Output Viterbi Algorithms
(SOVA) and Maximum A Posteriori (MAP) algorithms. The MAP algorithm is more efficient but more complex than the SOVA. However, simplified versions of this algorithm such as MAX-Log-MAP and Log-MAP perform almost as well with a reduced complexity.

\section{Proposed key reconciliation scheme}
\begin{figure}[h]
    \centering
    \includegraphics[width=9cm,height=2cm]{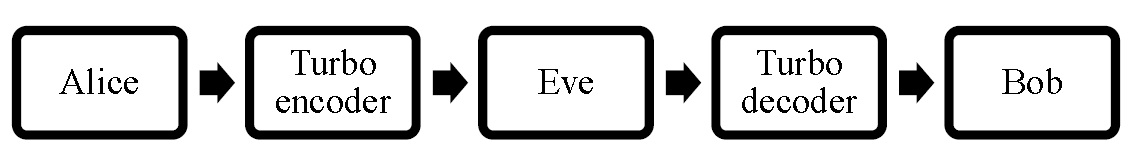}
 \caption{Proposed key reconciliation scheme.\label{Fig:method}}
\end{figure}

\begin{figure}[h]
    \centering
    \includegraphics[width=8cm,height=2cm]{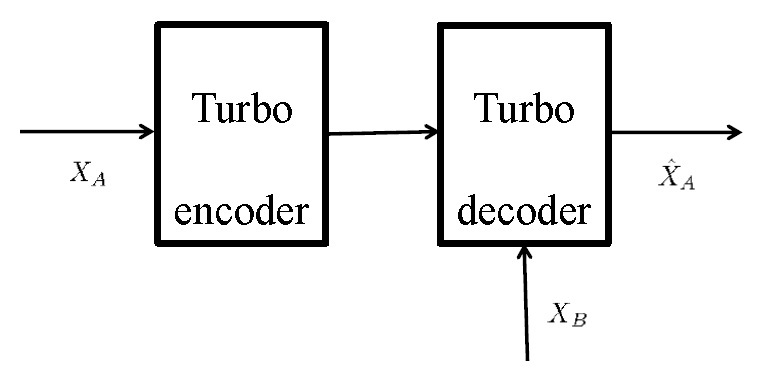}
 \caption{Source coding model with side information.\label{Fig:method_prim}}
\end{figure}

In our work, we propose to resort to Turbo codes to accomplish reconciliation, more precisely the analysis model
is shown in Fig.~\ref{Fig:method}. Alice and Bob communicate with BB84 protocol and Eve uses intercept and resend attack. The data sample generated by Alice is firstly encoded and then decoded in the Bob's side.
\\In what follows, we will only  operate on binary variables, which are applied to discrete variable Quantum Key Distribution. Let $X_{A}$  and $X_{B}$ be two of correlated variables belonging to Alice and Bob commonly called sifted keys. Reconciliation is a mechanism that allows them to eliminate the discrepancies between $X_{A}$ an $X_{B}$ and agree upon a string $\hat{X}_{A}$. Thus, the problem can be seen as a source coding problem with side information (see Fig.~\ref{Fig:method_prim})~\cite{rate}.

\section{Experimental results }
\begin{figure}[h]
    \centering
    \includegraphics[width=9cm,height=6cm]{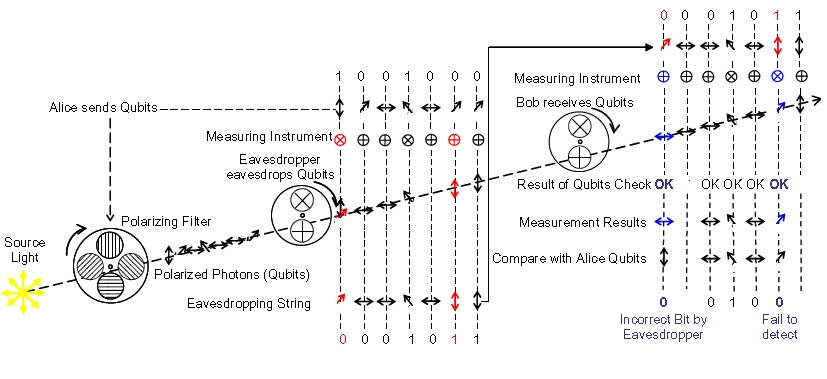}
    \caption{Intercept and Resend attack ($s$=1).\label{Fig:intercept}}
    \end{figure}

In this section, we discuss the experimental performances of our reconciliation method. We have implemented a Quantum Key Distribution protocol as described in~\cite{Bennett}. Also, we have implemented and experimented our reconciliation  method on the case of  one special type of eavesdropping strategy: Intercept and Resend.
\\Intercept and Resend is the most known eavesdropping individual strategy that can be implemented with actual technology means.
As described by Fig.~\ref{Fig:intercept}, Intercept and Resend attack is where Eve measures the quantum states sent by Alice. Then, she pretends as Alice and sends replacement states to Bob corresponding to her measurement result prepared in a chosen base from the two possible bases. This produces errors in the key Alice and Bob share. As Eve has no knowledge of the basis a state sent by Alice is encoded in, she can only guess which basis to measure in, in the same way as Bob. If she chooses correctly, she measures the correct photon polarization state as sent by Alice, and resents the correct state to Bob. However, if she chooses incorrectly the base, the state sent to Bob can't be the same as the one sent by Alice. Then, even if Bob measures this state in the Alice's basis, he gets a random result because Eve has sent him a state in the opposite basis.
 \\We define for this eavesdropping strategy the probability $s$ that one qubit sent by Alice to Bob is eavesdropped. The performance of the reconciliation method can be evaluated by measuring the Bit Error Rate and comparing it to the one generated by the Intercept and Resend attack given by~\cite{nedra}:
\begin{equation*}
BER=\frac{s}{4}.
\end{equation*}

\begin{figure}[!h]
    \centering
    \includegraphics[width=9cm,height=6cm]{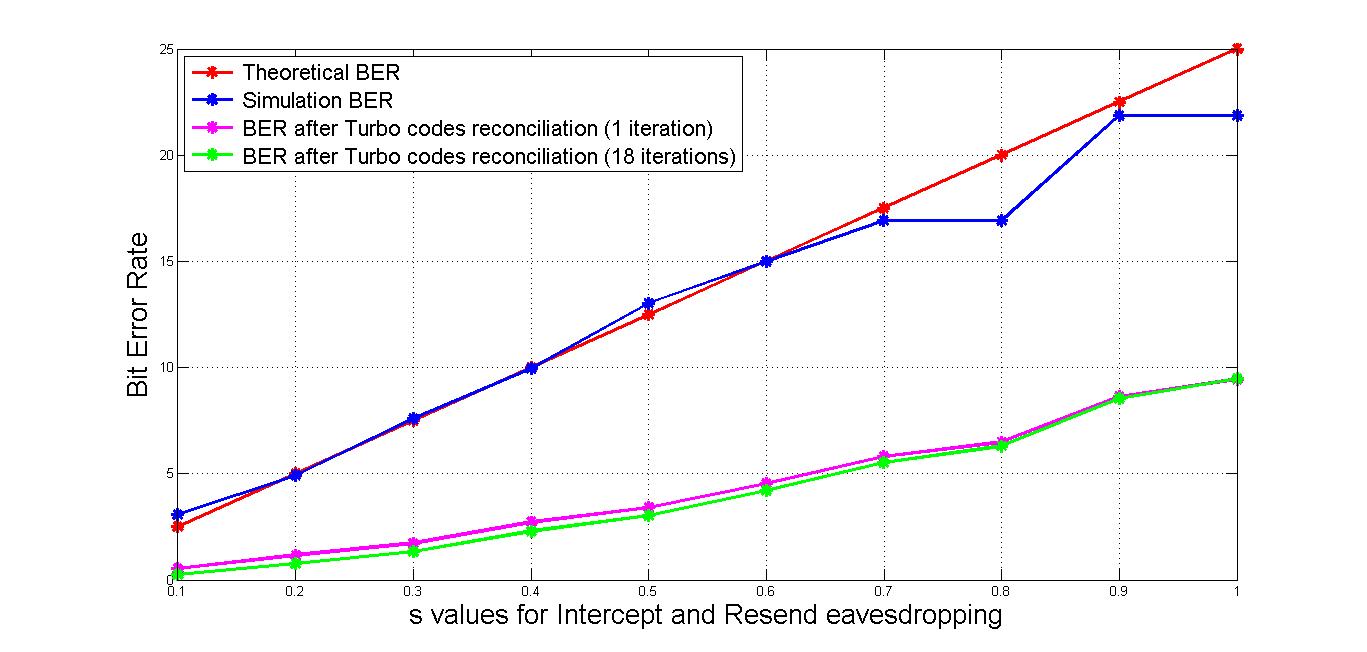}
    \caption{Bit Error Rate as function of $s$-values for Intercept and Resend eavesdropping.\label{Fig:prep}}
\end{figure}

 It can be noted that Bit Error Rate ($BER$) decreases  up to 0 as the parameter of eavesdropping attack  $s$ decreases also to 0. This means that when Eve has a little access to Alice states, Alice and Bob will have the same sifted keys because we have made the assumption that the channel is perfect. And, as the eavesdropper capability to catch Alice's states increases, we have an increasing Bit Error Rate as shown in Fig.~\ref{Fig:prep}.
\\In Fig.~\ref{Fig:prep}, the effect of the eavesdropping attack and the performance of our reconciliation method are also examined. The Bit Error Rate of the BB84 protocol decoupling with Turbo coding reconciliation scheme with 1 and 18 iterations is shown for different values of the eavesdropper parameter $s$. As a reference, we also show the theoretical and simulated Bit Error Rate without application of the reconciliation method.
Looking at the value of the parameter $s$ equal to 0.1, we see that the achieved Bit Error Rate amounts to $3.08\%$ for simulated BB84 protocol and decreases to $2.5\%$ thanks to the reconciliation, which means that it was devised by 1.23. The gain increases when the eavesdropper capability increases. Indeed, the achieved Bit Error Rate after reconciliation is $6.3\%$ for  $s$ equal to 0.8 and $16.92\%$ given by simulation of BB84 protocol without reconciliation, which means that it was devised by 5.12.
\begin{table}
\begin{centering}
\begin{tabular}{|c|c|c|}
\hline
 & {\scriptsize BB84 protocol} & {\scriptsize BB84 protocol with our reconciliation method}\tabularnewline
\hline
\hline
$s=0.1$ & $3.08\%
$ & $2.5\%
$\tabularnewline
\hline
$s=0.8$ & $16.92\%
$ & $6.3\%
$\tabularnewline
\hline
\end{tabular}
\par\end{centering}
\caption{Examples of Bit error Rate for Intercept and Resend attack on BB84 protocol. }
\end{table}

\begin{figure}[h]
    \centering
    \includegraphics[width=9cm,height=6cm]{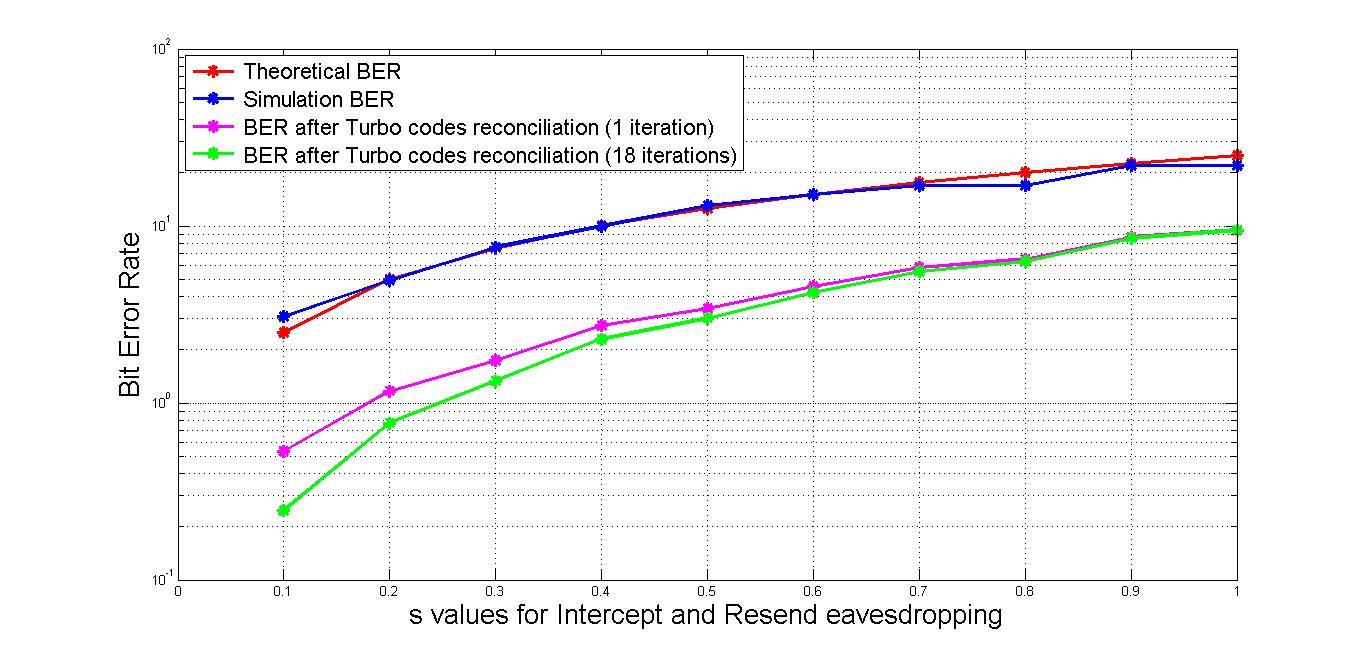}
    \caption{Bit Error Rate as function of $s$-values for Intercept and Resend eavesdropping in logarithmic domain .\label{Fig:result}}
\end{figure}

To sum up, we can mention that we have proven by simulation the efficiency of our reconciliation method as the average of initial Bit Error Rate equal to $13.11\%$ decreases to $4.17\%$ after reconciliation. Thus, our reconciliation method succeeds in removing on average of $68.19\%$ of the errors caused by eavesdropping.
One should note that the maximum Bit Error Rate admissible to distribute a secret is $11\%$ which is obtained with our method of reconciliation even with a total eavesdropping capability ($s=1$).
\\As it appears in Fig.~\ref{Fig:result}, the advantage of our reconciliation method can also be well seen in logarithmic presentation.
\\Two other criteria that one should keep in mind when evaluating a reconciliation method: its complexity and its
rapidity. This last criterium is especially relevant in the case of highly interactive schemes where latency can become an issue. We can say that the integration of our method of reconciliation doesn't affect so much the simulation time of the BB84 protocol. Just to have some order of magnitude, we give the BB84 protocol  time of running which is about 0.076s and the reconciliation method demands 0.023s.

Finally, we can see that at high value of the parameter $s$, the difference of BER gain resulting from an application of Turbo code with one and with 18 iterations is relatively small. So, it would be better to stop Turbo coding after few iterations since it doesn't affect the results and the number of iterations is of concern because every iteration reveals information and consumes times with each communication between Alice and Bob.
\section{Conclusion}
The BB84 protocol allows remote parties to share secret keys. But, the keys generated by this protocol will contain some errors which are caused by technical imperfections, as well as possibly by Eve's intervention. Such a situation, that the legitimate partners must remove the errors by public discussion called reconciliation. In this work, we suggest a way to improve the efficiency of the BB84 protocol by Turbo code reconciliation. We have shown that decoupling reconciliation and eavesdropping analysis in discreet variable quantum key  protocol by using Turbo codes allow remarkable improvement in the Bit Error Rate. There are many related works that are worth further investigation.
 We can generalize our research by taking into consideration not only eavesdropping effect, but also channel noise and other unwanted interactions in quantum computation and communication.

\end{document}